# A General Framework of Bound States in the Continuum in an Open Acoustic Resonator

**Lujun Huang[1#*], Bin Jia[2#], Artem S Pilipchuk[3#], Yankei Chiang[1], Sibo Huang[2], Junfei Li[4], Chen Shen[5], Evgeny N Bulgakov[3], Fu Deng[1], David A Powell[1], Steven A Cummer[4], Yong Li[2*], Almas F Sadreev[3*] and Andrey E Miroshnichenko[1*]**

[1] *School of Engineering and Information Technology, University of New South Wales, Canberra, Northcott Drive, ACT, 2600, Australia*

[2] *Institute of Acoustics, Tongji University, Shanghai, 200092, People's Republic of China*

[3] *L. V. Kirensky Institute of Physics, Krasnoyarsk 660036, Russia*

[4] *Department of Electrical and Computer Engineering, Duke University, Durham, North Carolina 27708, USA*

[5] *Department of Mechanical Engineering, Rowan University, Glassboro, NJ, 08028, USA*

\# These authors contributed equally to this work

[*] *ljhuang@mail.sitp.ac.cn, yongli@tongji.edu.cn, almas@tnp.krasn.ru, andrey.miroshnichenko@unsw.edu.au*

**Bound states in the continuum (BICs) provide a viable way of achieving high-Q resonances in both photonics and acoustics. In this work, we proposed a general method of constructing Friedrich-Wintgen (FW) BICs and accidental BICs in a coupled acoustic waveguide-resonator system. We demonstrated that FW BICs can be achieved with arbitrary two degenerate resonances in a closed resonator regardless of whether they have the same or opposite parity. Moreover, their eigenmode profiles can be arbitrarily engineered by adjusting the position of attached waveguide. That suggests an effective way of continuous switching the nature of BIC from FW BIC to symmetry-protected BIC or accidental BICs. Also, such BICs are sustained in the coupled waveguide-resonator system with shapes such as rectangle, ellipse, and rhomboid. These interesting phenomena are well explained by the two-level effective non Hermitian Hamiltonian, where two strongly coupled degenerate modes play a major role in forming such FW BICs. Besides, we found that such an open system also supports accidental BICs in geometry space instead of momentum space via tuning the position of attached waveguide, which are attributed to the quenched coupling between the waveguide and eigenmodes of the closed cavity. Finally, we fabricated a series of 3D coupled-resonator-waveguide and experimentally verified the existence of FW BICs and accidental BICs by measuring the transmission spectra. Our results complement the current BIC library in acoustics and provide new routes for designing novel acoustic devices, such as in acoustic absorbers, filters and sensors.**

# 1. Introduction

Bound states in the continuum (BICs) have triggered broad interests across the photonic community in the past ten years due to their extraordinary optical properties[1,2]. They corresponds to trapped modes despite being localized within the continuous spectrum, thus having an infinitely-large Q-factor. Usually, BICs must be converted into quasi-BICs (QBICs) with finite high-Q factors for practical application because only QBICs can be accessed by external excitation. The most salient property of BICs is that they suggest a viable way of realizing ultrahigh-Q resonances, whereas extreme field confinement is enabled for boosting light-matter interactions[3–6]. Different types of BICs or QBICs, including symmetry-protected (SP) BICs[7–10], accidental BICs[8,11], Friedrich-Wintgen (FW) BICs[12–16], and Fabry-Perot BICs[17–21], have been demonstrated in an array structure or single/few nanoparticles or photonic waveguide system.

In recent years, increasing attention has been paid to acoustic BICs, also recognized as trapped modes. SP BICs have been intensively studied by different groups[22–26]. For example, Evans et al considered SP BICs in a directional two-dimensional waveguide with a symmetrically loaded rigid circular obstacle[22]. Hein et al. found that placing an object in a duct can excite a quasi-trapped mode if symmetry is broken, leading to a high-Q Fano resonance[25]. Fabry-Perot BICs have been numerically studied by Hein et al.[27]. They found that Fabry-Perot BICs only happen when two resonators are separated by certain distances. FW BICs, known as embedded trapped modes, have been theoretically studied by Hein et al.[27], Lyapina et al.[28,29], and Dai et al[30].

Some early attempts at the experimental verification of SP BICs were conducted by Parker et al[31,32] and Cobelli et al[33]. Recently developed 3D printing technology enables researchers to verify these BICs. Huang et al. experimentally demonstrated the existence of SP BICs, FW BICs and mirror-induced BICs[34]. Sooner after, the same group demonstrated topological BICs[35] that arise from the merging of two Fabry-Perot BICs. Since the Q-factor of QBICs can be flexibly tuned by structure parameters, they have been used to realize perfect absorption for both acoustic wave[36] and elastic wave[37]. Although significant progress has been made in the past few years, there are still some open questions remaining unanswered: (1) Is it possible to construct an FW BIC based on any two degenerate resonances regardless of whether or not their parities are the same ? FW BICs have been found in both photonics and acoustics[13–15,28,34]. However, the two resonances must have the same parity and cross each other at certain geometry parameters such that their field profiles interchange with each other. (2) Can the eigenfield profile FW BICs be arbitrarily engineered ? Can the nature of BICs be switched from one type to another ? (3) Do accidental BICs exist in a finite system? Accidental BICs have been found at off-$\Gamma$ point in the first Brillouin zone in a photonic crystal slab[11]. However, the first Brillouin cannot be defined in such a finite system.

In this work, we reported a general frame work of BICs in an open acoustic system. We found that FW BICs can be constructed by any two degenerate resonances in a closed resonator no matter whether these two resonances have the same or opposite parity. Furthermore, by tuning the waveguide position, one can control the FW BIC's eigenfield profile arbitrarily, allowing for continuous transition from a FW BIC to a SP BIC (or an accidental BIC). Additionally, We found that such open systems support accidental BICs in geometry space by tuning the position of the attached waveguide. The formation mechanisms of both FW BICs and accidental BICs are explained by by the effective non Hermitian two-level Hamiltonian. Finally, we experimentally demonstrated the existence of FW BICs and accidental BICs by fabricating a series of 3D coupled resonator-waveguide samples and measuring their transmission spectra. The BICs are manifested by the vanished linewidth for collapse of the Fano resonance[38]. The Q-factor retrieved from the transmission spectrum is up to 340. Our findings on the new BICs may find intriguing applications in realizing high-performance acoustic devices, such as filters and sensors.

## 2. Results and discussion

### 2.1 General Friedrich-Wintgen BICs

We started by investigating the resonant modes in three types of coupled waveguide-resonator systems: single-port-waveguide-resonator system (Fig.1a-c) and two-port-waveguide-resonator systems with odd (Fig.1d-f) and even (Fig.1g-i) symmetry. For the sake of simplicity, we focus on the two-dimensional (2D) coupled waveguide-resonator systems to comprehensively illustrate all possible BICs in these three systems. Note that under certain circumstances BICs supported in Fig.1c are equivalent to BICs in Fig.1f and i, which is proved in section 2.3. Thus, we first consider the BICs in a single-port-waveguide-resonator system as shown in Fig.1c. Without loss of generality, we set the width of two waveguides as d=10cm and the height of rectangular resonator $L_y$=20cm. Also, we define the size ratio of the resonator as R=$L_y$/$L_x$, where $L_x$ is the width of rectangular resonator. In our previous study, we demonstrated that such an open non-Hermitian system supports a series of leaky modes, usually denoted as $M_{ml}$ (m and l are the number of antinodes along the x- and y-axis, respectively). Each leaky mode is represented by a complex eigenfrequency $\omega = \omega_0 - i\gamma$ with its Q-factor $Q = \frac{\omega_0}{2\gamma}$, where $\omega_0$ and $\gamma$ are respectively the resonant frequency and radiative decay rate. Importantly, the leaky modes supported by this open system play a dominant role in controlling the reflection (transmission) spectra, as demonstrated by temporal coupled-mode theory[34,39]. Thus, searching for BICs turns to finding the leaky modes with zero radiative decay rate or infinite Q-factor. Here, the leaky modes are calculated by commercial software package COMSOL Multiphysics.

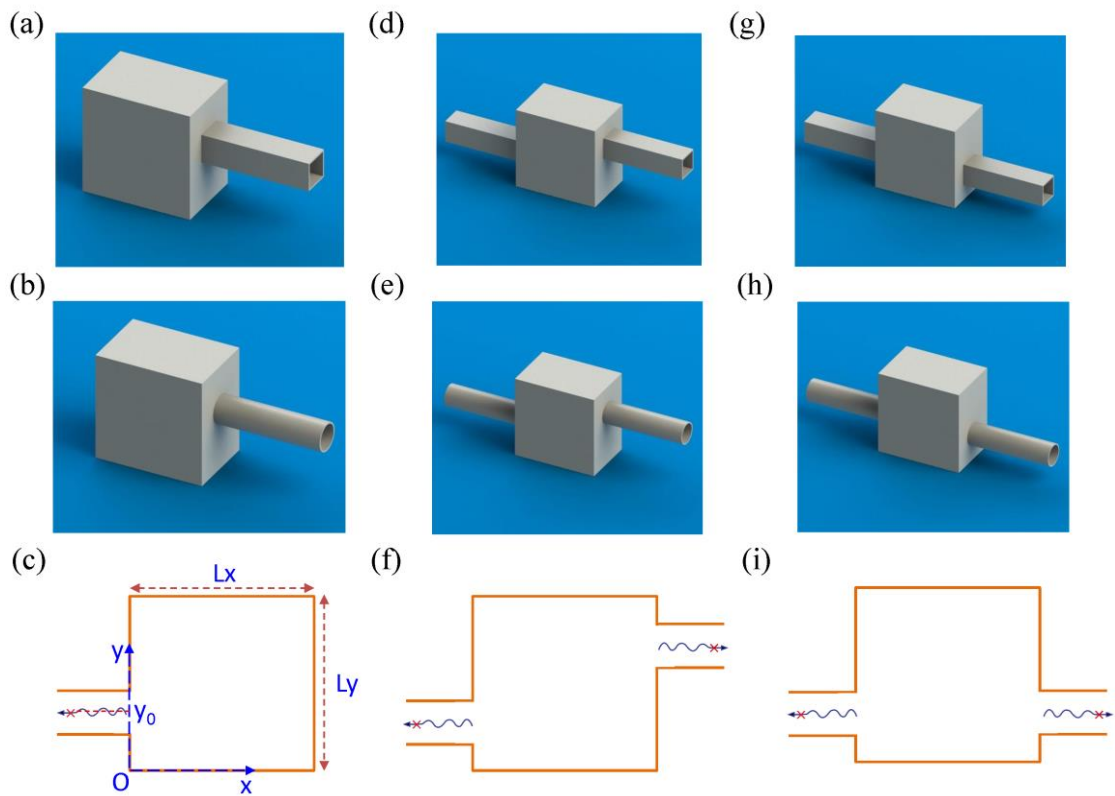

**Fig.1| Schematic drawing of an open resonator system**. a-c, Schematic illustration of a 3D single-port-waveguide resonator system with square or circular cross-section and their 2D equivalent. d-f, Schematic illustration of a 3D two-port-waveguide resonator system with odd symmetry and and their 2D equivalent. g-i, Schematic illustration of a 3D two-port-waveguide resonator system with even symmetry and and their 2D equivalent.

Remarkably, we find that the structure shown in Fig.1c supports many BICs. Fig.2a shows the lowest-order BIC with eigenfield distribution displayed in the inset when R varied from 0.95 to 1. Interestingly, this BIC also exists in the two-port system with odd-symmetry, as shown in Fig.2b. Careful examination on the eigenfield profile suggests that such a BIC is the superposition of eigenfield profile of modes $M_{12}$ and $M_{21}$ of a closed cavity (see Fig.2c), which can be fitted as

$$\psi_{BIC}(x,y) \approx 0.411 * \psi_{21}(x,y) - 0.912 * \psi_{12}(x,y) \qquad (1)$$

Where $\psi_{12}$ and $\psi_{21}$ represent the eigenfield distribution of closed cavity modes $M_{12}$ and $M_{21}$, respectively. We show how to derive the coefficients in the later section. It is worth noting that such a BIC has never been observed before, either in photonics or acoustics before. Interestingly, since the right and bottom hard boundaries can be viewed as two partial mirrors, this FW BIC can be correlated with the FW BICs in a full resonators sandwiched between two acoustic waveguides, as shown in Fig. 2d. The BIC's eigenfield profile is equivalent to one quarter of eigenfield profile of Friedrich-Wintgen BIC $M_{13}$ (see the inset figure of Fig.2d). More examples can be easily constructed by finding the FW BICs in a two-port-waveguide-full resonator system (See Fig.2e and Fig.S1). In the later section, we demonstrate that these BICs are FW BICs. For example, BIC in Fig.2a-b is the result of interference of eigenmodes $M_{12}$ and $M_{21}$ in a closed cavity although these two modes have opposite parity. A similar situation also occurs for the BIC in Fig. 2e, which is attributed to the interference of closed cavity modes $M_{31}$ and $M_{22}$. In addition, it is worth pointing out that a rectangular resonator is not the only resonator that can host such BICs. We also find this BIC in an elliptical resonator (Fig.2f) and a rhomboid resonator system (Fig.S2).

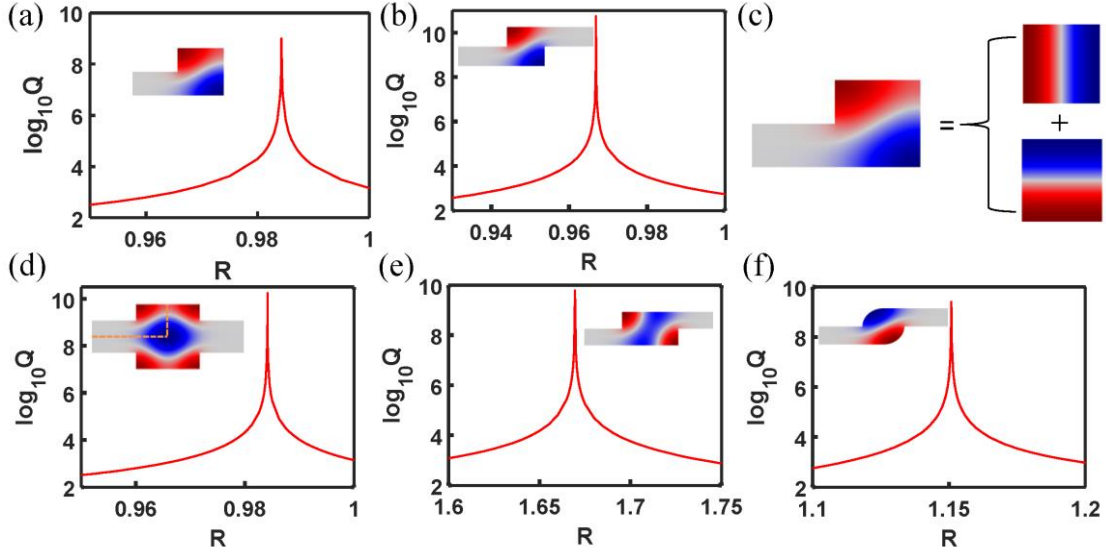

**Fig.2| FW BICs in an open resonator**. a, The Q-factor of an FW BIC in a single-port system vs the size ratio R. Inset shows the eigenfield profile of the BIC. b, The Q-factor of an FW BIC in a two-port system with odd-symmetry vs the size ratio R. Inset shows the eigenfield profile of BIC. c, Decomposition of an FW BIC into closed cavity modes $M_{12}$ and $M_{21}$. d, The Q-factor of an FW BIC in a coupled waveguide-full resonator system vs size ratio. e, The Q-factor of another FW BIC in a two-port-waveguide-resonator system with odd symmetry vs size ratio. f, The Q-factor of an FW BIC in an open elliptical resonator vs size ratio.

Note that such an FW BIC always exists even when the attached waveguide position is moved along four sides (Fig.S3a). Interestingly, its eigenfield profile can be arbitrarily engineered by adjusting the position of the attached waveguide, which is manifested in Fig.3a. Moreover, it is found that this FW BIC can be switched to a SP BIC when the attached waveguide position is moved from the bottom to the middle point. Any other intermediate state can be viewed as the superposition of two eigenmodes $M_{12}$ and $M_{21}$ of a closed cavity, which can be explicitly written as

$$\psi_{BIC}(x,y) \approx A * \psi_{21}(x,y) + B * \psi_{12}(x,y) \qquad (2)$$

Where A and B are the coefficients of the closed cavity modes $M_{21}$ and $M_{12}$, respectively. With accuracy of normalization we can choose $A = \cos\theta$ and $B = \sin\theta$. Then, BICs with eigenfields shown in Fig.3a can be reproduced by choosing an appropriate θ, where Fig.3b shows the synthetic eigenfield based on the eigenfield of $M_{21}$ and $M_{12}$ in the closed cavity. This may remind us of quantum mechanics, where a quantum state is usually decomposed into complete orthogonal eigenstates, and the absolute pre-coefficient squared indicates the possibility of finding the particle-like electrons. Importantly, similar phenomena also occur for other FW BICs, such as the one shown in Fig.2e (Fig.S3b and Fig.S4) and the one arising from destructive interference of closed cavity modes $M_{13}$ and $M_{31}$ (Fig.S3c and Fig.S5). More examples are included in the supporting information (Fig.S6-8). Thus, we can safely conclude that FW BICs always exist in systems shown in Fig.1c and f, and their eigenfields can be arbitrarily synthesized via tuning the position of the attached waveguide. Also, the nature of BICs can be switched from FW BICs to SP BICs.

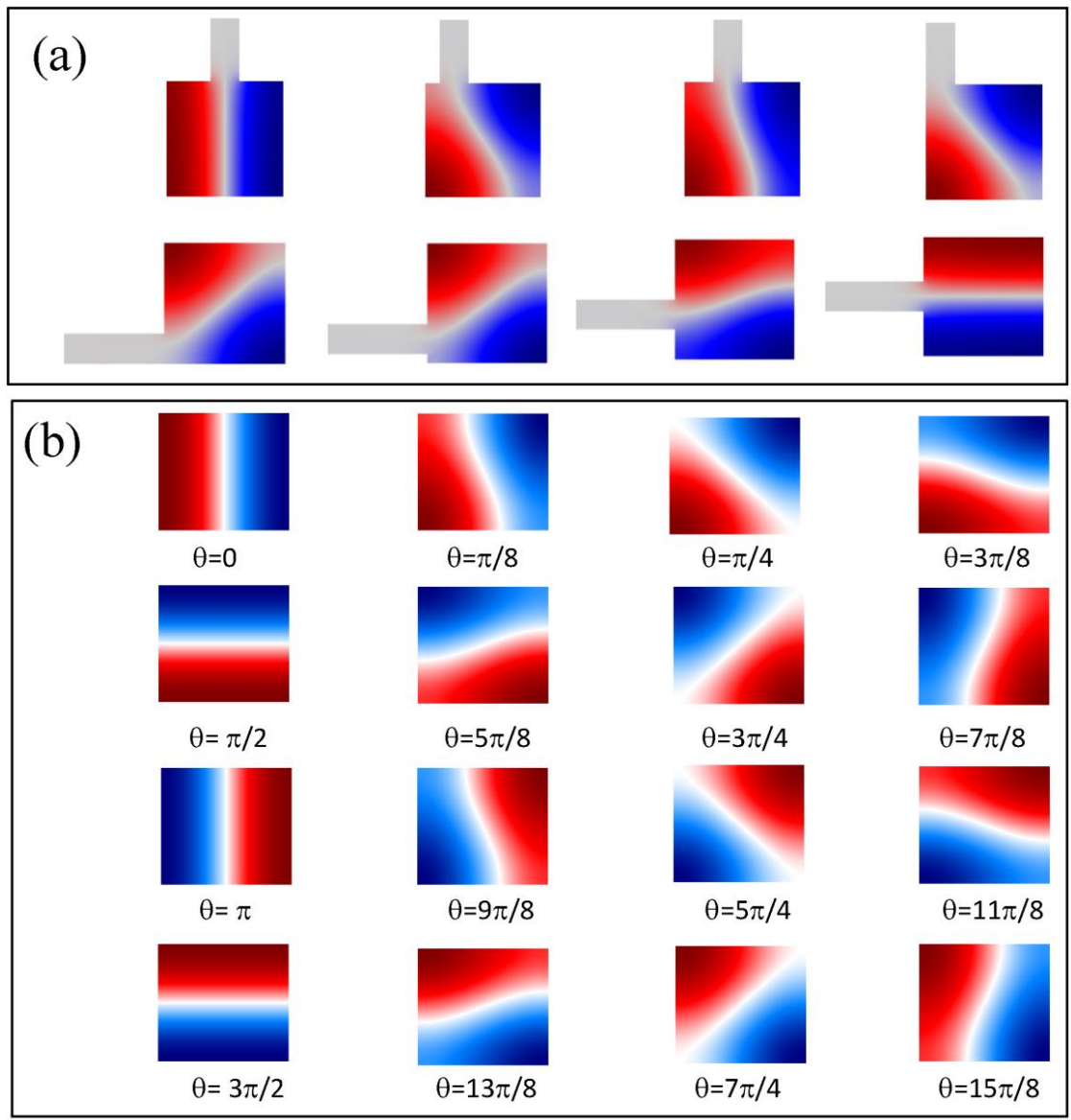

**Fig.3| Eigenfield profile engineering of FW BIC**. **a,** Eigenfield profile of a FW BIC for a single-port system with different attached waveguide positions. b, Synthetic eigenfield profile based on modes $M_{12}$ and $M_{21}$ in a closed square resonator.

**2.2 General Accidental BICs**

Accidental BICs were first demonstrated in a photonic crystal slab by Hsu et al[11]. They occur in the first Brillouin zone, located either in the x- and y-axis in momentum space or four quadrants depending on the structure parameters. However, it is difficult to define accidental BICs in a finite system using a similar strategy because one cannot define the first Brillouin zone in momentum space. Since accidental BICs in a periodic structure are found at off-Γ point while SP BICs are hosted at the Γ point, we may call BICs as accidental BICs in geometry space[40] when the attached waveguide deviates from the middle point that usually hosts SP BICs like $M_{12}$ and $M_{22}$. To illustrate this notion, we again consider the single-port-waveguide-resonator system. Here, the width of the waveguide is still set as d=10cm. The width and the height of the rectangular resonator are $L_x$=40cm and $L_y$=60cm, respectively. We have to point out here that the height and width of resonator can be any other values as long as the resonant frequencies of targeted mode are below the cut-off frequency of the waveguide. Fig.4a shows the Q-factor of mode $M_{13}$ as the center of the waveguide shifts from 5cm to 55cm. Surprisingly, two BICs occur at $y_{c1}$=15.45cm and $y_{c2}$=44.55cm, respectively. Fig.4b show the eigenfield distribution of two BICs. Note that these two BICs are not independent of each other. They follow $y_{c1}+y_{c2}=L_y$. This is because that mode $M_{13}$ is symmetric with respect to the yc=30cm. If there is a BIC at $y_{c0}$, another BIC appears at $y_c=L_y-y_{c0}$. Such BIC pairs are also found in a two-port-waveguide-resonator system with even or odd-symmetry, as confirmed in Fig.4c-f. Another common feature of these accidental BICs is that all critical positions are around $y_{c1}$=15cm and $y_{c2}$=45cm, which correspond to the nodes of closed cavity mode $M_{13}$. Similarly, one could construct such accidental BICs based on modes $M_{14}$ and $M_{23}$ by moving the attached waveguide position (Fig.S9).

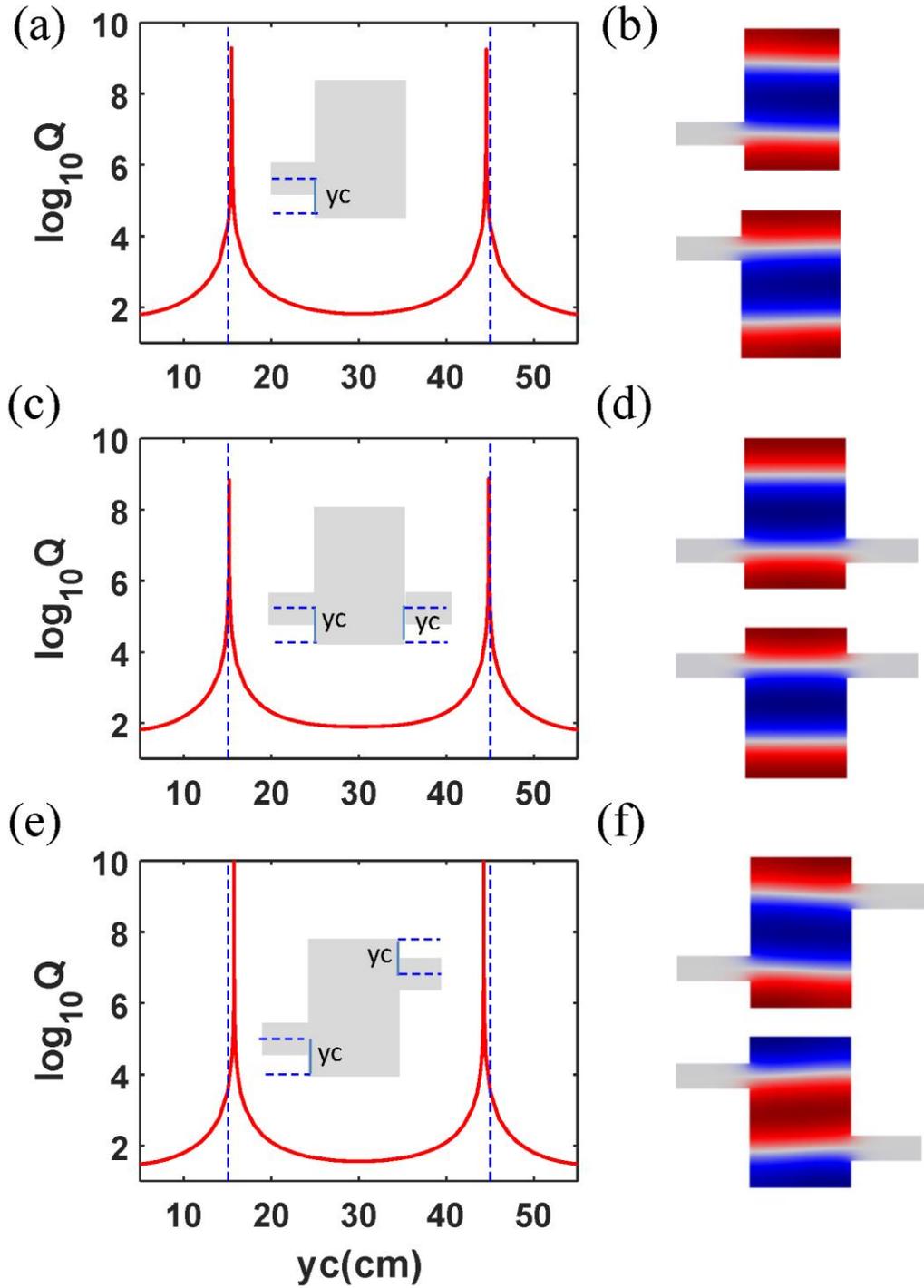

**Fig.4| Accidental BICs in an open resonator**. a, The Q-factor of accidental BICs in a single-port system with attached waveguide located at different position yc. b, Eigenfield profile of two accidental BICs in a single-port system. c, The Q-factor of accidental BICs in a two-port system with even-symmetry. d, Eigenfield profile of two accidental BICs in a two-port system with even-symmetry. e, The Q-factor of accidental BICs in a two-port system with odd-symmetry. f, Eigenfield profile of two accidental BICs in a two-port system with odd-symmetry. All three resonators have dimensions $L_x=40cm$ and $L_y=60cm$.

## 2.3 Formation mechanism of these two BICs

To account for these BICs' formation, we apply the method of an effective non-Hermitian Hamiltonian[2,41–45]. The Hamiltonian is the result of Feshbach projection of the total space onto the inner Hilbert space of eigenmodes in a closed resonator

$$H_{eff} = H_R - \sum_{p=1}^{\infty} \sum_C ik_p W_{Cp} W_{Cp}^+ \tag{3}$$

Where the first term $H_R$ represents the dynamics of the closed resonator, and the matrices $W_{cp}$ represent the coupling of resonators modes with the p-th channels of C-th waveguides with wavenumber $k_p$.

Again, we consider the coupled waveguide-resonator system shown in Fig.1c. To make the conclusion as general as possible, we set the width of waveguide d=1, and the width and height of the resonator are $L_x$ and $L_y$, respectively. Also, for the sake of convenience, we take the bottom left corner of the resonator as the origin, and the middle point of the attached waveguide position is $y = y_0$. Thus, the waveguide spans from $y = y_0 - 1/2$ to $y = y_0 + 1/2$. The first step is to compute the eigenfrequencies and eigenmodes of a closed resonator. They can be solved analytically with Neuman-boundary-conditions as follows,

$$\frac{v_{mn}^2}{\omega_0^2} = \frac{(m-1)^2}{L_x^2} + \frac{(n-1)^2}{L_y^2} \tag{4a}$$

$$\psi_{mn}(x,y) = \sqrt{\frac{(2-\delta_{m,1})(2-\delta_{n,1})}{L_x L_y}} \cos\left[\frac{\pi(m-1)x}{L_x}\right] \cos\left[\frac{\pi(n-1)y}{L_y}\right] \tag{4b}$$

Where $v_{mn}$ is the resonant frequency and $\omega_0 = \frac{\pi v}{d} = \pi v$, $v$ is the velocity of sound in the air.

The propagating wave numbers in the waveguide are given by

$$\frac{v^2}{\omega_0^2} = \frac{k_p^2}{\pi^2} + (p-1)^2 \tag{5a}$$

$$\phi_p(x,y) = \sqrt{(2-\delta_{p,1})} \cos[\pi(p-1)y] e^{ik_p x} \tag{5b}$$

Then the coupling matrix elements between eigenmodes of closed resonator and p-th propagation channels of the waveguide can be obtained by

$$W_{mn;p} = \int_{y_0-\frac{1}{2}}^{y_0+\frac{1}{2}} \psi_{m,n}(x=0,y)\phi_p(x=0,y)dy \tag{6}$$

After obtaining the coupling matrix, we calculate the complex eigenvalues of the effective Hamiltonian, whose real parts correspond to the resonant frequencies and imaginary parts are the half resonant linewidth. Thus, searching for BICs amounts to finding the zero imaginary part of eigenvalues. In general, the eigenfunction of each BIC can be decomposed as

$$\phi_{BIC}(x,y) = \sum_{mn} a_{mn}\psi_{mn}(x,y) \tag{7}$$

Since BIC is perfectly decoupled from the continuum, its eigenfunction must follow

$$\int_{y_0-\frac{1}{2}}^{y_0+\frac{1}{2}} \phi_{BIC}(x=0,y)dy = 0 \tag{8}$$

**2.3.1 Physical Mechanism of FW BICs**

When two resonant states approach each other as a function of a certain continuous parameter, interference causes an avoided crossing of the two states in their energy positions. Simultaneously, one of the resonant line widths vanishes exactly at a certain value of the parameters and the other is boosted to maximum. These are known as FW BIC[12]. They have been found in a single dielectric[13,14] and acoustic[28,34] resonator with rectangular cross-section. However, when constructing such FW BICs, two resonances must have the same parity and cross each other at a certain size ratio for the closed cavity. Thus, a pair of eigenmodes $M_{mn}$ and $M_{m+2, n-2}$ (or $M_{mn}$ and $M_{m-2, n+2}$) are frequently used for building FW BICs because they satisfy both requirements. Also, the eigenfield profile may interchange with each other when the size ratio of the rectangular resonator passes through the critical value.

Since the essence of finding FW BICs is to find two degenerate resonances in a closed resonator at a certain size ratio, in principle there should be numerous choices of paired modes, not limited to modes $M_{mn}$ and $M_{m+2, n-2}$. In the present paper, we consider the FW BIC in a rectangular resonator embedded into the first channel $p = 1$ provided that other channels are closed for $\nu < 1$. There are numerous degeneracies in a closed rectangular resonator

$$\frac{m^2}{L_x^2} + \frac{n^2}{L_y^2} = \frac{m'^2}{L_x^2} + \frac{n'^2}{L_y^2} \tag{9}$$

The lowest case corresponds to $m, n = 1,2$ and $m', n' = 2,1$ for square resonator $L_x = L_y$. The other degeneracies happening in a square resonator satisfy $m = n'$ and $n = m'$. The second-lowest degeneracy in a rectangular resonator is $m, n = 2,2$ and $m', n' = 3,1$ for $L_x = \sqrt{3}L_y$. All of these degeneracies can be used for constructing FW BICs in the single-port-waveguide-resonator system or two-port-waveguide-resonator system with odd-symmetry.

Fig. 5 shows the behavior of real and imaginary parts of complex eigenvalues for two resonances in the effective Hamiltonian. The solid line represents the results for the effective Hamiltonian considering 20 eigenstates in the closed resonator, and the dashed line corresponds to the case of effective Hamiltonian which only involves eigenmodes $M_{12}$ and $M_{21}$ of the closed resonator. From Fig.5, we indeed observe the avoided crossing at $L_x$=3.9942, around which one of the imaginary parts is suppressed to zero, and the other is boosted to maximum. These match the features of FW BICs. Interestingly, the formation of such a BIC can be mainly attributed to the destructive interference of modes $M_{12}$ and $M_{21}$ in a closed resonator, as demonstrated in the dashed line of Fig.5a-b. Thus, we may approximate the eigenfunction of the lowest-order FW BIC as a superposition of the two eigenmodes of the closed resonator and their coefficients A and B can be rigorously calculated by

$$\psi_{BIC}(x,y) \approx A * \psi_{21}(x,y) + B * \psi_{12}(x,y) \tag{10}$$

Substituting Eq.(10) into Eq.(8) gives us

$$A = W_{12;p=1} = \frac{1}{\pi}\sqrt{\frac{2L_y}{L_x}}\left[\sin\frac{\pi}{L_y}\left(y_0 + \frac{1}{2}\right) - \sin\frac{\pi}{L_y}\left(y_0 - \frac{1}{2}\right)\right] \tag{11a}$$

$$B = -W_{21;p=1} = -\sqrt{\frac{2}{L_x L_y}} \tag{11b}$$

Excellent agreement has been found between the eigenfield profile predicted from Eqs.(10-11) and numerical calculated eigenfield profile of FW BIC (Fig.S10) when $y_0$ varies from 0.5 to 2. Another interesting point is that such an FW-BIC is reduced to a SP BIC at $y_0 = \frac{L_y}{2}$ due to $W_{12;p=1} = 0$. Although modes $M_{12}$ and $M_{21}$ in the closed resonator play a major role in the formation of an FW BIC, the coupling between the evanescent modes of waveguide with imaginary $k_p$ ($p=2,3,…$) and eigenmodes in closed resonator leads to a slight deviation of $L_y$ from $L_y$=4. Following a similar strategy, we can construct more FW BICs by finding resonance degeneracies based on Eq. (7). Their formation can also be explained by the destructive interference between two major resonant modes. For example, the FW BIC shown in Fig.2e is the result of coupling between modes M₂₂ and M₃₁ (Fig.S11). The synthetic eigenfield profile of such an FW BIC is also confirmed using this analytical model (Fig.S12)

**2.3.2 Physical Mechanism of Accidental BICs and SP BICs**

When only the first channel p=1 is considered for ν<1, the key to realizing BICs is to find $W_{mn;p=1} = 0$. It is easy to find that zero coupling between waveguide and mode M$_{mn}$ happens at

$$y_0 = \frac{2s+1}{2(n-1)} L_y, s = 1,2,...,n-1 \qquad (12)$$

Eq.(12) predicts the critical waveguide positions around which BICs can be found in a real system. Note that accidental BICs only can be found for n>2. When n is even, both accidental BICs and SP BICs are supported in such a system. In Fig.4, dashed vertical lines indicate the predicted yc₀ with $W_{mn;p=1} = 0$, very close to the observed value yc of the BICs. The slight difference between them is ascribed to the fact that there is a small contribution of higher eigenmodes M₂₂ and M₂₃ owing to evanescent modes with p=2 (See section 1 in SI and Fig.S13).

**2.3.3 Symmetry Considerations**

As shown in section 2.1, the system shown in Fig.1f can support similar FW BICs to those shown in Fig.1c. For a two-port systems with even- or odd-symmetry, it is easy to derive

$$W^L_{mn;p=1} = (-1)^{m-1} W^R_{mn;p=1} \qquad (13a)$$

$$W^L_{mn;p=1} = (-1)^{m+n} W^R_{mn;p=1} \qquad (13b)$$

Thus, if FW BICs are sustained in a single-port-waveguide-resonator system, from Eq. (13b) one can immediately find such FW BICs in a two-port-waveguide-resonator system with odd-symmetry. Let's use the FW BIC shown in Fig.2a as an example to illustrate this notion, it has $C_1 W^L_{12;p=1} + C_2 W^L_{21;p=1} = 0$. According to Eq. (13b), odd structure symmetry requires $C_1 W^R_{12;p=1} + C_2 W^R_{21;p=1} = -C_1 W^L_{12;p=1} - C_2 W^L_{21;p=1} = 0$. Eq. (13) also explains why the lowest-order FW BIC in a square resonator is M₁₃ when the left and right waveguides are located at the middle point of left and right sides. Additionally, Eq. (13a) explains how two-port systems with either even- or odd-symmetry support accidental BICs like a single-port system.

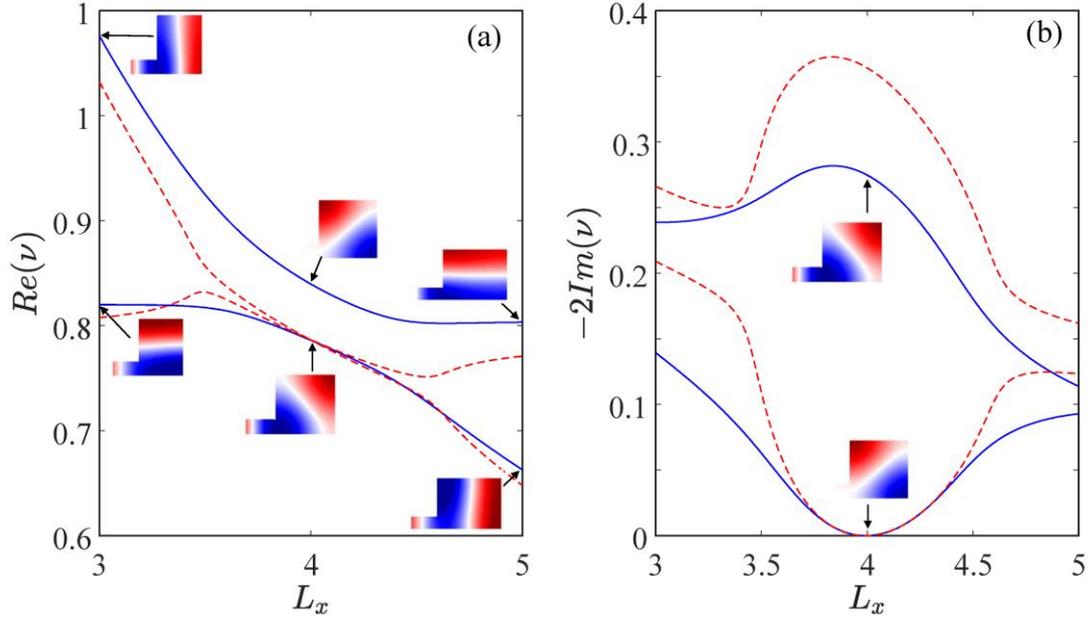

**Fig.5| Formation mechanism of FW BICs. a-b,** Real part (a) and imaginary part (b) of eigenfrequency in a single port system when 20 eigenstates are considered for calculating the complex eigenfrequencies (sold blue line). The dashed line corresponds to the case where only closed cavity modes $M_{12}$ and $M_{21}$ are considered.

## 2.4 Experimental verification of general FW BICs and accidental BICs

We would like to emphasize that such FW BICs and accidental BICs also exist in the 3D case. For experimental consideration, we study a 3D resonator coupled to two shifted cylindrical waveguides, as shown in Fig.6a. The diameter of the cylindrical waveguide is set as D=29mm. The resonator has dimensions $L_x$=2D=58mm, $L_y$=29mm. By sweeping $L_z$ from 49.7mm to 52.2mm, we find that there is a BIC at $L_z$=50.89mm, as confirmed in the top panel of Fig.6b. Its eigenfield distribution is shown on the right panel of Fig.6a, similar to the eigenfield profile of 2D case shown in Fig.2b. The emergence of BIC is also confirmed by calculating the transmission spectra mapping shown in top panel of Fig.6c, which indeed shows a vanishing linewidth at $L_z$=50.89mm. Then, we move to demonstrate the existence of such a BIC experimentally. We fabricate a series of samples with different $L_z$ covering the range from 49.7mm to 52.2mm. The bottom panel of Fig.6c shows the measured transmission spectra mapping. An excellent agreement is found between experiment and simulation. By applying a Fano-resonance fitting procedure[39,46] (Section S2 and Fig.S14), we retrieved the Q-factor of the acoustic resonance at different $L_z$. The relevant result is plotted in the bottom left panel of Fig.6b. The vertical dashed line indicates the $L_z$ showing the vanished linewidth of resonance, indicating the formation of the BIC. The experimentally measured Q-factor shows several orders of magnitude reduction compared to the theoretical prediction because of viscous losses in the real system. However, we still observed a relatively high-Q resonance with Q-factor up to 340, which is large enough for real applications. Besides, we emphasize that this is not the only structure supporting such an FW BIC. More examples can be found in supporting information (Fig.S15).

We also demonstrate the accidental BICs in such a two-port system. Fig.6d shows the optical image of 3D-printed sample. The diameters of the two waveguides are also D=29mm. The dimensions of the cuboid resonator are $L_x$=70mm, $L_y$=36mm and $L_z$=50mm. By varying the attached waveguide position, we find an accidental BIC at xc=19.11mm, as seen in the top panel of Fig.6e. This critical position deviates slightly from the $xc_0=L_x/4$=17.5mm predicted by Eq. (11) because of the coupling between the evanescent modes in the waveguide and eigenmodes in the closed resonator. The existence of such a BIC is confirmed by the calculated transmission spectra mapping shown in the top panel of Fig.6f. We fabricated nine samples using 3D-printing technology (See Materials and Methods section for details). Excellent agreements can be found between the measured transmission spectra (bottom panel of Fig.6f) and simulated transmission spectra. We also retrieve their Q-factor by Fano-fitting and plotted them in the bottom panel of Fig.6e. The overall trend of Q-factor matches reasonably well to the calculated Q-factor although the measured one is several orders lower than the numerical prediction due to the viscous losses inside the waveguide and resonator. However, all Q-factors are above 200, large enough for many applications. Also, we find further examples of accidental BICs in such a two-port-waveguide-3D resonator-system (Fig.S16).

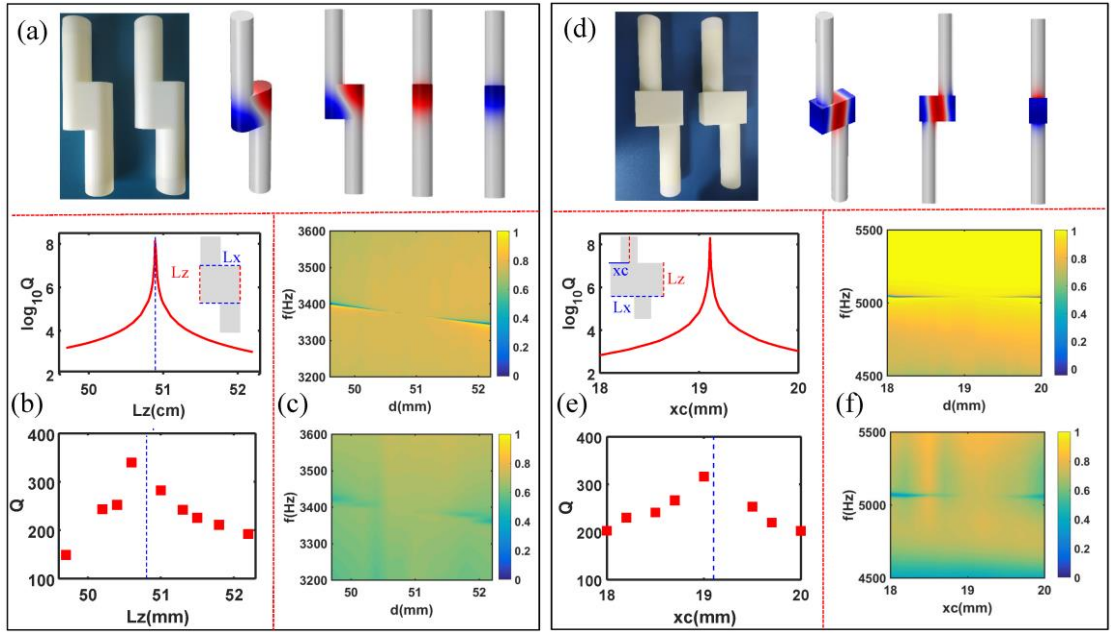

**Fig.6| Experimental demonstration of an FW BIC and an accidental BIC in a 3D open resonator with odd symmetry**. **a,** The left panel is the optical image of the 3D printed sample and the right panel represents the eigenfield profile of an FW BIC. **b,** Calculated and measured Q-factor vs Lz, **c,** Calculated and measured transmission spectra mapping for Lz varying from 49.7mm to 52.2mm. **d,** The left panel is the optical image of the 3D printed sample and right panel represents the eigenfield profile of an accidental BIC. **e,** Calculated and measured Q-factor vs xc, **f,** Calculated and measured transmission spectra mapping for xc varying from 18mm to 20mm.

## 3. Conclusion

In summary, we reported a general procedure for finding FW BICs and accidental BICs in an open acoustic resonator. We demonstrated that FW BICs can be constructed by any two degenerate resonances with either the same or opposite parity in a single-port system or two-port system with odd-symmetry. Moreover, their eigenfield profile can be arbitrarily synthesized through tuning the waveguide position, similar to the case of quantum mechanics. Also, FW BICs can be switched to either SP BICs or accidental BICs when waveguide is placed at a certain position. In addition, we found that such an open system supports accidental BICs in geometry space when its symmetry is broken. The formation of both FW BICs and accidental BICs is well explained by the effective Hamiltonian method. The former is attributed to the destructive interference of two eigenmodes in the closed cavity, while the latter arises from the suppressed coupling between propagation modes in the waveguide and eigenmode in the closed resonator. Finally, we experimentally demonstrated both FW BICs and accidental BICs in a 3D coupled-waveguide-resonator system with odd symmetry. The emergence of BICs is evidenced by the vanished line width of the acoustic resonance. Our findings promise many exciting applications, such as enhanced acoustic emission, ultra-narrowband acoustic absorbers, acoustic filters, and sensors.

## Materials and Methods

### Simulations

All simulations are performed with the commercial software package COMSOL Multiphysics. The speed of sound and density of air are 343 m/s and 1.29kg/m$^3$, respectively. When calculating the eigenmodes and transmission (or reflection spectrum), we apply perfect matched layer boundaries at the two ends of waveguides to mimic acoustic wave propagation in the infinite space. The other exterior boundaries are set as rigid.

### Experiments

The experimental samples are fabricated by 3D-printing technology using laser sintering stereolithography (SLA, 140μm) with a photosensitive resin (UV curable resin), exhibiting a manufacturing precision of 0.1 mm. The complex transmission (and reflection) coefficients of the samples are measured using a Brüel & Kjær type-4206T impedance tube with a diameter of 29 mm. A loudspeaker generates a plane wave, and the amplitude and phase of local pressure are measured by four 1/4-inch condenser microphones (Brüel & Kjær type-4187) situated at designated positions. The complex transmission (and reflection) coefficients are obtained by the transfer matrix method.

## Acknowledgements

L. Huang and A. E. Miroshnichenko were supported by the Australian Research Council Discovery Project (DP200101353) and the UNSW Scientia Fellowship program. Y. K. Chiang and D. A. Powell were supported by the Australian Research Council Discovery Project (DP200101708), B. Jia, S. Huang, and Y. Li were supported by the National Natural Science Foundation of China (Grants No. 12074286), and Shanghai Science and Technology Committee under grant No. 21JC1405600. A. Pilipchuk and A. Sadreev were supported by Russian Science Foundation with Grant number 22-12-00070.

## Author Contributions

L. Huang and A. E. Miroshnichenko conceived the idea. L. Huang, A. S. Pilipchuk and A. F. Sadreev performed the theoretical calculation and numerical simulation. B. Jia, S. Huang, and Y. Li fabricated the sample and performed the reflection/transmission spectra measurements. Y. K. Chiang, J. Li, C. Shen, E. N. Bulgakov, F. Deng, D. A. Powell and S. Cummer helped with the numerical simulation. L. Huang, Y. Li, A. F. Sadreev, and A. E. Miroshnichenko supervised the project. All authors discussed the results and prepared the manuscript.

## References

1. Hsu, C. W., Zhen, B., Stone, A. D., Joannopoulos, J. D. & Soljačić, M. Bound states in the continuum. *Nat. Rev. Mater.* **1**, 16048 (2016).

2. Sadreev, A. F. Interference traps waves in an open system: bound states in the continuum. *Reports Prog. Phys.* **84**, 55901 (2021).

3. Kodigala, A. *et al.* Lasing action from photonic bound states in continuum. *Nature* **541**, 196 (2017).

4. Xu, L. *et al.* Dynamic Nonlinear Image Tuning through Magnetic Dipole Quasi-BIC Ultrathin Resonators. *Adv. Sci.* **6**, 1802119 (2019).

5. Volkovskaya, I. *et al.* Multipolar second-harmonic generation from high-Q quasi-BIC states in subwavelength resonators. *Nanophotonics* **9**, 3953–3963 (2020).

6. Koshelev, K. *et al.* Nonlinear Metasurfaces Governed by Bound States in the Continuum. *ACS Photonics* **6**, 1639–1644 (2019).

7. Koshelev, K., Lepeshov, S., Liu, M., Bogdanov, A. & Kivshar, Y. Asymmetric Metasurfaces with High-$Q$ Resonances Governed by Bound States in the Continuum. *Phys. Rev. Lett.* **121**, 193903 (2018).

8. Bulgakov, E. N. & Sadreev, A. F. Bloch bound states in the radiation continuum in a periodic array of dielectric rods. *Phys. Rev. A* **90**, 53801 (2014).

9. Lee, J. *et al.* Observation and Differentiation of Unique High-$Q$ Optical Resonances Near Zero Wave

Vector in Macroscopic Photonic Crystal Slabs. *Phys. Rev. Lett.* **109**, 67401 (2012).

10. Li, S., Zhou, C., Liu, T. & Xiao, S. Symmetry-protected bound states in the continuum supported by all-dielectric metasurfaces. *Phys. Rev. A* **100**, 63803 (2019).

11. Hsu, C. W. *et al.* Observation of trapped light within the radiation continuum. *Nature* **499**, 188–191 (2013).

12. Friedrich, H. & Wintgen, D. Interfering resonances and bound states in the continuum. *Phys. Rev. A* **32**, 3231–3242 (1985).

13. Huang, L., Xu, L., Rahmani, M., Neshev, D. & Miroshnichenko, A. E. Pushing the limit of high-Q mode of a single dielectric nanocavity. *Adv. Photonics* **3**, (2021).

14. Rybin, M. V. *et al.* High- Q Supercavity Modes in Subwavelength Dielectric Resonators. *Phys. Rev. Lett.* **119**, 1–5 (2017).

15. Sadreev, A. F., Bulgakov, E. N. & Rotter, I. Bound states in the continuum in open quantum billiards with a variable shape. *Phys. Rev. B* **73**, 235342 (2006).

16. Lepetit, T. & Kanté, B. Controlling multipolar radiation with symmetries for electromagnetic bound states in the continuum. *Phys. Rev. B* **90**, 241103 (2014).

17. Sato, Y. *et al.* Strong coupling between distant photonic nanocavities and its dynamic control. *Nat. Photonics* **6**, 56–61 (2012).

18. Suh, W., Yanik, M. F., Solgaard, O. & Fan, S. Displacement-sensitive photonic crystal structures based on guided resonance in photonic crystal slabs. *Appl. Phys. Lett.* **82**, 1999–2001 (2003).

19. Bulgakov, E. N. & Sadreev, A. F. Bound states in the continuum in photonic waveguides inspired by defects. *Phys. Rev. B* **78**, 75105 (2008).

20. Weimann, S. *et al.* Compact Surface Fano States Embedded in the Continuum of Waveguide Arrays. *Phys. Rev. Lett.* **111**, 240403 (2013).

21. Marinica, D. C., Borisov, A. G. & Shabanov, S. V. Bound states in the continuum in photonics. *Phys. Rev. Lett.* **100**, 1–4 (2008).

22. Evans, D. & Porter, R. Trapped modes embedded in the continuous spectrum. *Q. J. Mech. Appl. Math.* **51**, 263–274 (1998).

23. Linton, C. M., McIver, M., McIver, P., Ratcliffe, K. & Zhang, J. Trapped modes for off-centre structures in guides. *Wave Motion* **36**, 67–85 (2002).

24. Hein, S. & Koch, W. Acoustic resonances and trapped modes in pipes and tunnels. *J. Fluid Mech.* **605**, 401–428 (2008).

25. HEIN, S., KOCH, W. & NANNEN, L. Fano resonances in acoustics. *J. Fluid Mech.* **664**, 238–264 (2010).

26. DUAN, Y., KOCH, W., LINTON, C. M. & McIVER, M. Complex resonances and trapped modes in ducted domains. *J. Fluid Mech.* **571**, 119–147 (2007).

27. Hein, S., Koch, W. & Nannen, L. Trapped modes and Fano resonances in two-dimensional acoustical duct-cavity systems. *J. Fluid Mech.* **692**, 257–287 (2012).

28. Lyapina, A. A., Maksimov, D. N., Pilipchuk, A. S. & Sadreev, A. F. Bound states in the continuum in

open acoustic resonators. *J. Fluid Mech.* **780**, 370–387 (2015).

29. Lyapina, A. A., Pilipchuk, A. S. & Sadreev, A. F. Trapped modes in a non-axisymmetric cylindrical waveguide. *J. Sound Vib.* **421**, 48–60 (2018).

30. Dai, X. Total Reflection of Two Guided Waves for Embedded Trapped Modes. *AIAA J.* **59**, 131–139 (2020).

31. Parker, R. Resonance effects in wake shedding from parallel plates: Some experimental observations. *J. Sound Vib.* **4**, 62–72 (1966).

32. Parker, R. Resonance effects in wake shedding from parallel plates: Calculation of resonant frequencies. *J. Sound Vib.* **5**, 330–343 (1967).

33. Cobelli, P. J., Pagneux, V., Maurel, A. & Petitjeans, P. Experimental observation of trapped modes in a water wave channel. *{EPL} (Europhysics Lett.* **88**, 20006 (2009).

34. Huang, L. *et al.* Sound Trapping in an Open Resonator. *Nat. Commun.* **12**, 4819 (2021).

35. Huang, L. *et al.* Topological Supercavity Resonances in the Finite System. *Adv. Sci.* **n/a**, 2200257 (2022).

36. Huang, S. *et al.* Extreme Sound Confinement From Quasibound States in the Continuum. *Phys. Rev. Appl.* **14**, 21001 (2020).

37. Cao, L. *et al.* Perfect absorption of flexural waves induced by bound state in the continuum. *Extrem. Mech. Lett.* **47**, 101364 (2021).

38. Kim, C. S., Satanin, A. M., Joe, Y. S. & Cosby, R. M. Resonant tunneling in a quantum waveguide: Effect of a finite-size attractive impurity. *Phys. Rev. B* **60**, 10962–10970 (1999).

39. Fan, S., Suh, W. & Joannopoulos, J. D. Temporal coupled-mode theory for the Fano resonance in optical resonators. *J. Opt. Soc. Am. A* **20**, 569–572 (2003).

40. Pilipchuk, A. S. & Sadreev, A. F. Accidental bound states in the continuum in an open Sinai billiard. *Phys. Lett. A* **381**, 720–724 (2017).

41. Feshbach, H. Unified theory of nuclear reactions. *Ann. Phys. (N. Y).* **5**, 357–390 (1958).

42. Dittes, F.-M. The decay of quantum systems with a small number of open channels. *Phys. Rep.* **339**, 215–316 (2000).

43. Okołowicz, J., Płoszajczak, M. & Rotter, I. Dynamics of quantum systems embedded in a continuum. *Phys. Rep.* **374**, 271–383 (2003).

44. Sadreev, A. F. & Rotter, I. S-matrix theory for transmission through billiards in tight-binding approach. *J. Phys. A. Math. Gen.* **36**, 11413–11433 (2003).

45. Maksimov, D. N., Sadreev, A. F., Lyapina, A. A. & Pilipchuk, A. S. Coupled mode theory for acoustic resonators. *Wave Motion* **56**, 52–66 (2015).

46. Miroshnichenko, A. E., Flach, S. & Kivshar, Y. S. Fano resonances in nanoscale structures. *Rev. Mod. Phys.* **82**, 2257–2298 (2010).